\documentclass[showpacs,twocolumn,amssymb,prl,aps]{revtex4}
\usepackage{graphicx,epsfig}

\newcommand{\etal}{\textit{~et~al.}}
\begin{document}
\title{Knudsen Diffusion in Silicon Nanochannels}
\author{Simon~Gruener and Patrick Huber}
\email[E-mail: ]{p.huber@physik.uni-saarland.de}
\affiliation{Faculty of Physics and Mechatronics Engineering, Saarland University, D-66041 Saarbr\"ucken, Germany}


\begin{abstract}
Measurements on helium and argon gas flow through an array of parallel, linear channels of $12$~nm diameter and $200$~$\rm\mu$m length in a single crystalline silicon membrane reveal a Knudsen diffusion type transport from $10^{\rm 2}$ to $10^{\rm 7}$ in Knudsen number Kn. The classic scaling prediction for the transport diffusion coefficient on temperature and mass of diffusing species, $D_{\rm He}\propto \sqrt{T}$ is confirmed over a $T$-range from 40~K to 300~K for He and for the ratio of $D_{\rm He}/D_{\rm Ar} \propto \sqrt{m_{\rm Ar}/m_{\rm He}}$. Deviations of the channels from a cylindrical form, resolved with electron microscopy down to subnanometer scales, quantitatively account for a reduced diffusivity as compared to Knudsen diffusion in ideal tubular channels. The membrane permeation experiments are described over 10 orders of magnitude in Kn, encompassing the transition flow regime, by the unified flow model of Beskok and Karniadakis.
\end{abstract}


\maketitle
Knudsen diffusion (KD) refers to a gas transport regime where the mean free path between particle-particle collisions $\lambda$ is significantly larger than at least one characteristic spatial dimension $d$ of the system considered \cite{Knudsen1909}. By virtue of the negligible mutual particle collisions transport in such systems takes place via a series of free flights and statistical flight direction changes after collisions with the confining walls. Because of the dependency of $\lambda$ on pressure $p$ and temperature $T$ given by the kinetic theory of gases, $\lambda \propto T/p$, this transport regime is only observable in macroscopic systems at very low $p$ or elevated $T$. By contrast, for transport in spatially nanoconfined systems with $d=\mathcal{O}(10$ nm$)$, the Knudsen number $Kn=\lambda/d$, which quantifies the gas rarefaction, is larger than 1 even for ambient pressures and temperatures, e.g. $\lambda(\text{He})=118$ nm at $p=1$~bar and $T=297$~K. Thus, for many processes involving gas transport in restricted geometries, e.g. gas catalysis and storage\cite{Relev} or equilibration phenomena via gas flow in meso- and nanopores\cite{Wallacher2004,PEq}, KD plays a crucial role.\\
\begin{figure}[htbp]
\epsfig{file=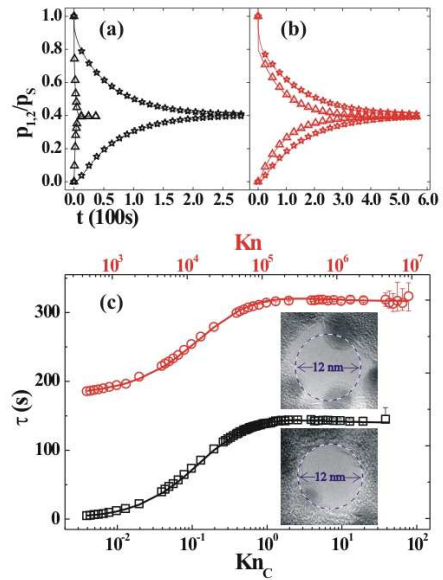, angle=0, width=0.75\columnwidth} \caption{\label{pRelaxEmptyMembrane}(color online) He pressure relaxations in R1 and R2 at $T=297$~K for starting pressures {$p_{\rm s}=0.1$ mbar} (stars) and {$100$ mbar} (triangles) without (a) and with SiNC membrane (b). (c) Pressure relaxation time $\tau$ as determined from measurements without (squares) and with SiNC membrane (circles) versus Knudsen number in the capillaries $Kn_{\rm c}$ (bottom axis) and in the SiNCs $Kn$ (top axis). The lines in (a), (b) and (c) represent calculated $p(t)$ and $\tau(Kn)$ values. Insets in (c): TEM cross section views of two SiNCs in comparison with circular perimeters of 12~nm diameter (dashed lines).}
\end{figure}
Albeit the phenomenon of KD has been known for almost a hundred years, details on its microscopic mechanisms are still at debate, in particular how wall roughness down to atomic scales and particle-wall interaction parameters affect the statistics of the diffusion process and thus the value of the KD transport coefficient $D$ \cite{Theo}. In the past, a comparison between theory and experiment has often been hampered by complex pore network topologies, tortuous transport paths and poorly characterized pore wall structures in available nanoporous matrices. Nowadays, membranes with better defined geometries, such as carbon nanotube bundles\cite{Hinds2004, Holt2006}, porous alumina \cite{Roy2003}, and tailored pores in mesoporous silicon \cite{Wallacher2004, Striemer2007}, allow us to address such fundamental questions in more detail. A better understanding of, and more in-depth information on, transport in such systems is not only of academic interest, but also of importance for the architecture and functionality in the emerging field of nanoelectromechanical and nanofluidic systems \cite{Karniadakis2005, Whitby2007}.

In this Letter, we report the first rigorous experimental study on transport of simple rare gases, i.e. He and Ar through an array of parallel aligned, linear channels with $\sim12$~nm diameter and $200$~$\mu$m length in single crystalline Si over a wide temperature ($40 \text{ K}<T<300$ K) and $Kn$-range ($10^{\rm 2}<Kn<10^{\rm 7}$). We explore the classical scaling predictions given by Knudsen for $D$ on $T$, mass of diffusing species, $m$, and explore its dependency on $Kn$. Attention shall also be paid to the morphology of the channel walls, resolved with transmission electron microscopy (TEM), and how it affects the diffusion.

Linear, non-interconnected channels oriented along the $\langle100\rangle$ Si crystallographic direction (SiNCs) are electrochemically etched into the surface of a Si (100) wafer \cite{Lehmann1991}. After the channels reach the desired length $L$ of $200\pm 5~\mu$m, the anodization current is increased by a factor of 10 with the result that the SiNC array is detached from the bulk wafer \cite{Henschel2007}. The crystalline Si walls call for irregular channel perimeters as one can see in the TEM pictures of Fig.~\ref{pRelaxEmptyMembrane}(c), recorded from a membrane part Ar ion milled to $\sim5$~nm thickness. In first approximation, however, the SiNC's cross-section can be described as circular with a diameter of $\sim12$~nm, in accordance with the analysis of an Ar sorption experiment at $T=86$~K, which yields additionally a SiNCs density of $\Phi = (1.5\pm 0.1) \cdot 10^{\rm 11}$ cm$^{\rm -2}$.

Our experimental setup consists of a copper cell with an inlet and outlet opening \cite{MFA2007}. Inlet and outlet are connected via stainless steel capillaries of radius $r_{\rm c}=0.7$~mm and length $l_{\rm c}=70$~cm with two gas reservoirs, R1 and R2, of an all-metal gas handling. Additionally two pneumatic valves V1 and V2 are used to open and close the connections between the sample cell and R1 and the sample cell and R2. Four thermostatted capacitive pressure gauges allow us to measure the gas pressures in R1 and R2, $p_{\rm 1}$ and $p_{\rm 2}$, resp., over a wide pressure range (5$\cdot 10^{\rm -3}$~mbar $<p<$1~bar) with an accuracy of $10^{\rm -3}$~mbar. The cell is mounted in a closed-cycle He cryostat to control the temperature between 40~K and 300~K with an accuracy of $1$~mK.

Our goal of studying gas transport dynamics over a large $Kn$ range necessitates a thorough understanding of the intrinsic flow characteristics of our apparatus. Therefore, we start with measurements on He flow through the macroscopic capillaries and the empty sample cell at room temperature $T=297$~K. We record the equilibration of $p_{\rm 1}(t)$ and $p_{\rm 2}(t)$ towards a pressure $p_{\rm eq}$ as a function of time $t$ after initial conditions, $p_{\rm 1}=p_{\rm s}>p_{\rm 2}=0$~mbar at $t=0$~s. In Fig.~\ref{pRelaxEmptyMembrane}~(a) $p_{\rm 1}(t)/p_{\rm s}$ and $p_{\rm 2}(t)/p_{\rm s}$ are plotted for starting pressures of $p_{\rm s}=0.1$ and 100~mbar. We observe pressure equilibrations towards $p_{\rm eq}={0.4 \cdot p_{\rm s}}$. The value 0.4 is dictated by the volume ratio of R1 to R2. From the $p(t)$-curves we derive characteristic relaxation times $\tau$ according to the recipe $p_{\rm 1}(t=\tau)-p_{\rm 2}(t=\tau)\equiv 1/10 \cdot p_{\rm s}$. It is understood that $p$ is monotonically decreasing downstream from R1 to R2. In order to quantify the gas rarefaction we resort, therefore, to a calculation of a mean Knudsen number $Kn_{\rm c}=\lambda(\overline{p})/r_{\rm c}$ in the capillaries assuming a mean pressure, $\overline{p}=p_{\rm eq}$. This simplification is justified by the analysis provided below and by theoretical studies which indicate differences of less than 1\% between flow rates calculated with an exact and an averaged treatment for $Kn$ \cite{Sharipov1994, Karniadakis2005}. In Fig.~\ref{pRelaxEmptyMembrane}~(c) we plot measured $\tau$ values versus $Kn_{\rm c}$ corresponding to a $p_{\rm s}$ variation from $0.01$ to $100$~mbar. For $Kn_{\rm c}<0.1$, $\tau$ increases with increasing $Kn_{\rm c}$. In an intermediate $Kn_{\rm c}$ range, $0.1<Kn_{\rm c}<0.6$, we observe a cross-over regime towards a saturation plateau with $\tau\sim140$~s that extends to the largest $Kn_{\rm c}$ studied. These changes in the $p$ relaxation and hence flow dynamics are reminiscent of the three distinct transport regimes known to occur for gases as a function of their rarefaction \cite{Karniadakis2005, Tabeling2005}: For $Kn_{\rm c}<0.1$ the number of interparticle collisions still predominates over the number of particle-wall collisions. Hagen-Poiseuille's law is valid and predicts a decreasing flow rate and hence increasing $\tau$ due to the $1/Kn$-scaling of the particle number density in gas flows. For $Kn_{\rm c}>1$ we enter the pure KD regime\cite{Knudsen1909}, where theory predicts a He KD transport coefficient $D_{\rm c}^{\rm He}$ dependent, however a $Kn_{\rm c}$ independent particle flow rate,
\begin{equation}
\label{nKD}
\dot n_{\rm Kn} = \pi \frac{r_{\rm c}^2}{l_{\rm c}} \frac{p_{\rm o}-p_{\rm i}}{k_{\rm B} T} D_{\rm c}^{\rm He} ,
\end{equation}
which is responsible for the plateau in $\tau$ for the larger $Kn_{\rm c}$ investigated. In Eq.~\ref{nKD} $p_{\rm i}$, $p_{\rm o}$ refers to the inlet and outlet pressure of the capillary considered, resp., and $k_{\rm B}$ to the Boltzmann factor. In the intermediate $Kn$-range, the interparticle collisions occur as often as particle-wall collisions which gives rise to the cross-over behavior found for $\tau$.

We elucidate this behavior in more detail by dividing the flow path within our apparatus into two flow segments (up- and downstream capillary) and calculate the particle number changes along the flow path and the resulting $p_{\rm 1}(t)$, $p_{\rm 2}(t)$ with a $1$~ms resolution using the \textit{unified flow model} of Beskok and Karniadakis (BK-model)\cite{Karniadakis2005, Beskok1999, Gruener2006} and a local $Kn$-number for each flow segment, $Kn_{\rm l}=Kn((p_{\rm i}+p_{\rm o})/2)$:
\begin{equation}
\label{nBK}
\dot n_{\rm BK} = \frac{\pi r_{\rm c}^4 (p_{\rm o}^{\rm 2}-p_{\rm i}^{\rm 2})}{16\, l_{\rm c}\, k_{\rm B}\, T}  \frac{1+\alpha\,Kn_{\rm l}}{\mu(T)} \left (1+\frac{4 \, Kn_{\rm l}}{1-b \, \, Kn_{\rm l}} \right ).
\end{equation}
Eq.~\ref{nBK} comprises a Hagen-Poiseuille term, a term which treats the transition of the transport coefficient from continuum-like, i.e. the bare dynamic viscosity $\mu(T)$, to the KD transport coefficient, $D_{\rm c}^{\rm He}$ with $\alpha=2.716/\pi \tan ^{\rm -1} (\alpha_{\rm 1} Kn^{\beta})$, and a generalized velocity slip term, which is second-order accurate in $Kn$ in the slip and early transition flow regimes ($Kn<0.5$). The model captures for $Kn\rightarrow0$ the no-slip Hagen-Poiseuille limit, whereas it transforms to Eq.~\ref{nKD} for $Kn \gg 1$. As verified by comparison of the BK-model with Direct Simulation Monte Carlo and solutions of the Boltzmann equation, a choice of $b=-1$ for the slip parameter results in the correct velocity profile and flowrate, as well as a proper pressure and shear stress distribution in a wide $Kn$-range, including the transition flow regime. After optimizing the free parameters in Eq.~\ref{nBK}, $\alpha_{\rm 1}$ and $\beta$, we arrive at the $p(t)$- and $\tau(Kn_{\rm c})$-curves depicted in Fig.~\ref{pRelaxEmptyMembrane}~(a) and (c), resp. The good agreement of the BK-model predictions with our measurements is evident and the extracted parameters $\alpha_{\rm 1}=10\pm0.2$ and $\beta=0.5\pm0.05$ ($D_{\rm c}^{\rm He}=0.412\pm0.06$~m$^{\rm 2}$/s) agree with BK-modelling of He gas flow as a function of rarefaction \cite{Karniadakis2005, Tison1993}. An analogous analysis for Ar gas flow yields $\alpha_{\rm 1}=1.55\pm0.05$ and $\beta=0.1\pm0.02$ ($D_{\rm c}^{\rm Ar}=0.134\pm0.02$~m$^{\rm 2}$/s).

\begin{figure}[htbp]
\epsfig{file=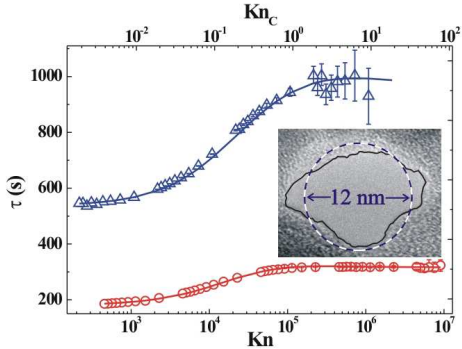, angle=0, width=0.8\columnwidth}\caption{\label{pSiArHeDiff}(color online) Pressure relaxation time $\tau$ for He (circles) and Ar (triangles) measured with built-in SiNC membrane at $T=297$~K versus Knudsen number in the SiNCs $Kn$ (bottom axis) and in the capillaries $Kn_{\rm c}$ (top axis). The lines represent calculated $\tau(Kn)$-curves. Inset: TEM cross section of a SiNC in comparison with a circular channel perimeter of 12~nm diameter (dashed line). The SiNC's perimeter is highlighted by a line.}
\end{figure}
Having achieved a detailed understanding of the intrinsic flow characteristics of our apparatus, we can now turn to measurements with the SiNC membrane. The membrane is epoxy-sealed in a copper ring \cite{MFA2007} and special attention is paid to a careful determination of the accessible membrane area, $A=0.79\pm 0.016$~cm$^{\rm 2}$ in order to allow for a reliable determination of the number of SiNCs inserted into the flow path, $N=\Phi \cdot A =(11.8\pm 1)\cdot 10^{\rm 10}$. As expected the pressure equilibration is significantly slowed down after installing the membrane - compare panel (a) and (b) in Fig.~\ref{pRelaxEmptyMembrane}. Choosing selected $p_{\rm s}$ within the range 0.005~mbar to 100~mbar, which corresponds to a variation of $Kn$ in the SiNCs from $10^{\rm 2}$ to $10^{\rm 7}$, we find an increase in $\tau$ of $\sim170$~s, when compared to the empty cell measurements - see Fig.~\ref{pRelaxEmptyMembrane}~(c).
From the bare offset in $\tau$ for $Kn_{\rm c}>1$, we could calculate $D_{\rm He}$ in the SiNCs. We are, however, interested in the $D_{\rm He}$ behavior in a wide $Kn$-range, therefore we modify our flow model by inserting a segment with KD transport mechanism characteristic of $N$ tubular channels with $d=12$~nm and $L=200$~$\rm\mu$m in between the capillary flow segments. Adjusting the single free parameter in our simulation, the value of the He diffusion coefficient $D_{\rm He}$ in a single SiNC, we arrive at the $p(t)$- and $\tau(Kn)$-curves presented as solid lines in Fig.~\ref{pRelaxEmptyMembrane}~(b) and (c), resp. The agreement with our measurement is excellent and the model yields a $Kn$ independent $D_{\rm He}=3.76\pm 0.8$~mm$^{\rm 2}$/s.

It is worthwhile to compare this value with Knudsen's prediction for $D$. In his seminal paper he derives an expression for $D$ with two contributions, a factor $G$, characteristic of the KD effectivity of the channel's shape, and a factor solely determined by the mean thermal velocity of the particles $\overline{v}$ given by the kinetic theory of gases,
\begin{equation}
\label{gD}
D = \frac{1}{3}\;G\;\overline{v} = \frac{1}{3}\; G \; \sqrt{\frac{8 k_{\rm B} T}{\pi m}}.
\end{equation}
Interestingly, by an analysis of the number of particles crossing a given section of a channel in unit time after completely diffuse reflection from an arbitrary element of wall surface and while assuming an equal collision accessibility of all surface elements, Knudsen derived an analytical expression for $G$ for a channel of \textit{arbitrary} shape. Knudsen's second assumption is, however, only strictly valid for circular channel cross-sections, as pointed out by v. Smoluchowski \cite{Knudsen1909} and elaborated for fractal pore wall morphologies by Coppens and Froment \cite{Theo}. Given the roughly circular SiNCs' cross section shapes, we nevertheless resort to Knudsen's formula, here quoted normalized to the KD form factor for a perfect cylinder:
\begin{equation}
\label{gf}
\frac{1}{G} = \frac{1}{4 L}\;\int _0 ^L { \frac{o(l)}{A(l)}\; dl}.
\end{equation}
The integral in Eq.~\ref{gf} depends on the ratio of perimeter length $o(l)$ and cross sectional area $A(l)$ along the channel's long axis direction $l$, only. This ratio is optimized by a circle, accordingly the most effective KD channel shape is a cylinder, provided one calls for a fixed $A$ along the channel. Eq.~\ref{gf} yields due to our normalization just $G=d$ and we would expect a $D_{\rm He}$ of $5$~mm$^{\rm 2}$/s for He KD in a SiNC, if it were a cylindrical channel of $d=12$~nm. A value which is $\sim 30\%$ larger than our measured one. If one recalls the TEM pictures, which clearly indicate non-circular cross-sections this finding is not surprising. In fact these pictures deliver precisely the information needed for an estimation of the influence of the SiNC's irregularities on the KD dynamics. We determine the ratio $o/A$ of 20 SiNC cross sections, and therefrom values of $G$. The values of the SiNC in Fig.~\ref{pRelaxEmptyMembrane}~(top), (bottom) and Fig.~\ref{pSiArHeDiff} corresponds to $G$s of 9.1~nm, 10.3~nm, and 10.6~nm, resp. Tacitly assuming that the irregularities exhibited on the perimeter are of similar type as along the channel's long axis we take an ensemble average over the 20 $G$ values and arrive at a mean $G$ of 9.9~nm, which yields a $D_{\rm He}$ of $4.1$~mm$^{\rm 2}$/s, a value which agrees within the error margins with our measured one. 

\begin{figure}[htbp]
\epsfig{file=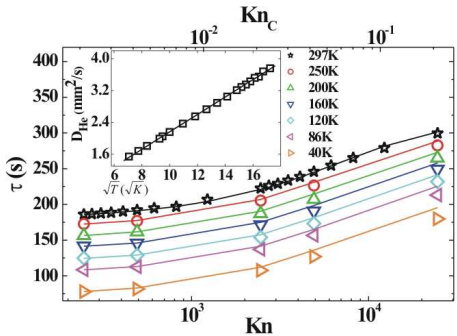, angle=0, width=0.8\columnwidth}
\caption{\label{pSiKnDiffTemp} (color online) Pressure relaxation time $\tau$ for He gas flow through SiNC membrane (symbols) and simulated $\tau$(Kn) (lines) versus Knudsen number in the SiNC membrane $Kn$ (bottom axis) and in the supply capillaries $Kn_{\rm c}$ (top axis) at selected temperatures $T$. Inset: He diffusion coefficient $D_{\rm He}$ in SiNCs in comparison with the $\sqrt{T}$ prediction of Eq.~\ref{gD} (line) plotted versus $\sqrt{T}$.}
\end{figure}

To further explore the KD transport dynamics in the SiNCs we now focus on the $m$ and $T$ dependency of $D$. In Fig.~\ref{pSiArHeDiff} the $\tau(Kn)$ curve recorded for Ar and He at $T=297$~K are presented. Both exhibit a similar form, the one of Ar is, however, shifted up markedly towards larger $\tau$. Our computer model can quantitatively account for this slowed-down dynamics by a factor $2.97\pm0.25$ decrease in $D$ as compared to the He measurements, which confirms the prediction of Eq.~\ref{gD}, $D_{\rm He }/D_{\rm Ar}=\sqrt{m_{\rm Ar}/m_{\rm He}}=3.16$. For the exploration of the $T$ behavior we choose again He as working fluid due to its negligible physisorption tendency, even at low $T$. We perform measurements at selected $T$s from $297$ K down to $40$ K, shown in Fig.~\ref{pSiKnDiffTemp}. We again perform computer calculations assuming KD in the SiNC array superimposed to the transport in the supply capillaries, presented as lines in Fig.~\ref{pSiKnDiffTemp}, and optimize the single free parameter $D(T)$ in order to match the observed $\tau(Kn,T)$ behavior. Despite unresolvable deviations at larger $Kn_{\rm c}$ and decreasing $T$, which we attribute to thermal creep effects, characteristic of $T$ gradients along the flow path \cite{Karniadakis2005}, we find, in agreement with the experiment, increasingly faster dynamics with decreasing $T$. Note, this counter-intuitive finding for a diffusion process results from the $1/T$ scaling in Eq.~\ref{nKD}, which reflects the $T$-dependency of the particle number density in gas flows. By contrast, the $D_{\rm He}(T)$, determined by our simulations, scales in excellent agreement with $\sqrt{T}$, see inset in Fig.~\ref{pSiKnDiffTemp}, confirming Knudsen's classic result down to the lowest $T$ investigated.

We find no hints of anomalous fast KD here, as was recently reported for a bundle of linear carbon nanotubes \cite{Holt2006} and explained by an highly increased fraction of specularly reflected particles upon wall collisions \cite{Knudsen1909}. The altered collision statistics was attributed to the crystalline structure and the atomical smoothness of the nanotube walls. The SiNC walls are also crystalline, however, not atomical flat, as can be seen from our TEM analysis. Along with the formation of a native oxide layer, typical of Si surfaces, which renders the near surface structure amorphous, silica like, this, presumably, accounts for the normal KD observed here contrasting the one found for the graphitic walls of carbon nanotubes.

We presented here the first detailed study of gas transport in linear SiNCs. Our conclusions are drawn from a correct treatment of gas flow over 10 orders of magnitude in gas rarefaction, which is, to the best of our knowledge, the largest $Kn$ range ever explored experimentally. The characteristic properties of KD, an independency of $D$ on $Kn$, its scaling predictions on $m$, $T$, and on details of the channel's structure, resolved with sub-nm resolution, are clearly exhibited by our measurements.

\begin{acknowledgments}
We thank A.~Beskok for helpful discussions and the German Research Foundation (DFG) for support within the priority program 1164, \textit{Nano- \& Microfluidics} (Hu 850/2).
\end{acknowledgments}



\end{document}